# Doping of Ce in $T$-$La_2CuO_4$: Rigorous test for electron-hole symmetry for high-$T_c$ superconductivity

A. Tsukada [1, 2], H. Yamamoto [3], M. Naito [1]

[1] Department of Applied Physics, Tokyo University of Agriculture and Technology, 2-24-16 Naka-cho, Koganei, Tokyo 184-8588, Japan

[2] NTT Basic Research Laboratories, NTT Corporation, 3-1 Morinosato-Wakamiya, Atsugi, Kanagawa 243-0198, Japan

[3] NTT Science and Core Technology Laboratory Group, NTT Corporation, 3-1 Morinosato-Wakamiya, Atsugi, Kanagawa 243-0198, Japan

We report that Ce doping was achieved in $La_2CuO_4$ with the $K_2NiF_4$ ($T$) structure for the first time by molecular beam epitaxy. A synthesis temperature of as low as ~ 630ºC and an appropriate substrate choice, *i.e.*, (001)$LaSrGaO_4$ ($a_s$ = 3.843 Å), enabled us to incorporate Ce into the $K_2NiF_4$ lattice and to obtain Ce-doped $T$-$La_{2-x}Ce_xCuO_4$ up to $x$ ~ 0.06. The doping of Ce makes $T$-$La_2CuO_4$ more insulating, which is in sharp contrast to Sr (or Ba) doping in $T$-$La_2CuO_4$, which makes the compound metallic and superconducting. The observed smooth increase in resistivity from the hole-doped side ($T$-$La_{2-x}Sr_xCuO_4$) to the electron-doped side ($T$-$La_{2-x}Ce_xCuO_4$) indicates that the electron-hole symmetry is broken in the $T$-phase materials.

## I. INTRODUCTION

The playground for high-$T_c$ superconductivity is the $CuO_2$ plane common to both *p*- and *n*-type high-$T_c$ superconductivity, and the electronic phase diagram of high-$T_c$ cuprates is roughly symmetric between *p*- and *n*-type doping.[1] Hence, it has been claimed that "electron-hole" symmetry holds for high-$T_c$ superconductivity. Based on this claim, the doped Mott insulator scenario has been widely accepted,[2] in which the parent material is a Mott insulator (anti-ferromagnetic insulator) and high-$T_c$ superconductivity develops when the insulator is exposed to either *p*- or *n*-type doping. However, it should be borne in mind that electron-hole symmetry is far from obvious and even surprising. Since the mother compounds of high-$T_c$ superconductors can be regarded as charge-transfer insulators, doped holes go on the oxygen sites but doped electrons go on the Cu sites, which should result in doped carriers with quite different natures. Furthermore it must be emphasized that the argument for the above "electron-hole" symmetry is based on a comparison of *p*- and *n*-type doping in different structures, namely, hole doping in the $K_2NiF_4$ (*T*) structure [*e.g.*, $La_{2-x}Sr_xCuO_4$ (LSCO)] (Ref. 3) and electron doping in the $Nd_2CuO_4$ (*T'*) structure (*e.g.*, $Nd_{2-x}Ce_xCuO_4$).[4]

In principle, it is desirable to compare hole and electron doping in the same crystal structure. However, such a comparison has not yet been undertaken because it is empirically known in bulk synthesis that hole doping is possible only in octahedral ($CuO_6$) or pyramidal ($CuO_5$) cuprates whereas electron doping is possible only in square-planar ($CuO_4$) cuprates. For example, electron doping in the *T* structure or hole doping in the *T'* structure has never been achieved in bulk synthesis. However, in this article we report that Ce can be incorporated into the *T* lattice [*T*-$La_2CuO_4$ (LCO)] by employing a low-temperature synthetic route with molecular beam epitaxy (MBE), as evident from a monotonic change in the *c*-axis lattice constant. The solubility limit of Ce in the *T* lattice depends on the substrate material, and can be extended up to $x \sim 0.06$ with (001)$LaSrGaO_4$. This made it possible to perform a rigorous test on the "electron-hole" symmetry in the *T*-structured compounds. The result revealed that Ce doping makes *T*-LCO more insulating, which is in sharp contrast to Sr (or Ba) doping in *T*-LCO, which makes the compound metallic and superconducting. The observed smooth increase in resistivity from the hole-doped side (*T*-LSCO) to the electron-doped side [*T*-$La_{2-x}Ce_xCuO_4$ (LCCO)] indicates that the electron-hole symmetry is broken in the *T*-phase materials.

## II. EXPERIMENT

*T*-LCCO thin films were grown in a custom-designed MBE chamber (base pressure ~ $10^{-9}$ Torr) from metal sources using multiple electron-gun evaporators with atomic oxygen (1 sccm) activated by an RF power of 300 W. The details of our MBE growth process are described elsewhere.[5] There are two problems specific to the growth of *T*-LCCO films. The first is that Ce tends to segregate out from the *T* lattice at growth temperatures ($T_s$) higher than 700˚C.[6] Films grown at $T_s$ > 700˚C showed no change in the *c*-axis lattice constant even when the amount of Ce was changed, indicating that Ce is not incorporated into the lattice. This problem forced us to lower the growth temperature to well below 700˚C, which led to the second problem, namely the inclusion of *T'*-LCCO as an impurity phase. At low synthesis temperatures, the *T'* phase is easy to form and even predominates in the preparation of LCCO films even for *x* values as small as 0.05.[7, 8] *T'*-LCCO is three to six orders of magnitude more conductive than *T*-LCCO and even exhibits superconductivity. Therefore, just a tiny inclusion of *T'* materials ruins the generic transport properties of the *T*-phase materials. As a trade-off between the first and second factors, we employed $T_s$ = 630°C.

Epitaxial stabilization is another way to overcome the above competing problems. As we demonstrated in a previous work on the phase control of undoped $La_2CuO_4$,[9] substrate materials have a strong influence on the selective stabilization of the *T* versus *T'* structure. Table I lists the substrate materials used in the present work. Our previous results indicated that the in-plane lattice constant ($a_s$) and the crystal structures of the substrate materials are crucial with regard to selective phase stabilization: substrates with $a_s$ close to 3.80 Å and with the $K_2NiF_4$ structure have a strong tendency toward *T*-phase stabilization.[9] Utilizing epitaxial stabilization, we attempted to maximize the solubility limit of Ce in *T*-$La_2CuO_4$. The typical film thickness was designed to be only 450 Å in order to make full use of the epitaxial stabilization.

After growth, most films were cooled to ambient temperature in a vacuum with $P_{O2}$ of less than $10^{-8}$ Torr to avoid the introduction of excess oxygen into the films. For comparison, some films were cooled in an ozone atmosphere of $P_{O3}$ at $10^{-5}$ Torr to introduce excess oxygen into the films. The films were characterized by X-ray diffraction (XRD) and resistivity measurements. All the Ce-doped *T*-phase films had a very high resistance (typically > 100 MΩ) even at 300 K, so electrodes with a low contact resistance were required for reliable measurements. We formed the electrodes by Ag evaporation. The sampling time for the resistivity measurements also had to be set at several seconds. The valence state of Ce was determined by X-ray photoelectron spectroscopy (XPS) measurements, in which the films were transferred to a surface analysis chamber in a vacuum via a gate valve.

## III. RESULTS AND DISCUSSION

### A. Ce doping in the *T* lattice

Figure 1 shows the XRD patterns of LCCO films grown on $LaAlO_3$ (LAO) substrates with different Ce concentrations (*x*). The XRD patterns indicate that the films are *c*-axis oriented. Since the *c*-axis lattice constants ($c_0^f$) of *T*- and *T'*-LCCO are clearly different, the phase identification is rather straightforward.[9] The calculated XRD patterns are also included in Fig. 1 for reference. The films on the LAO substrates are single-phase *T* for $x \leq 0.045$ and single-phase *T'* for $x \geq 0.105$. The films are a two-phase mixture of *T* and *T'* for *x* = 0.06 - 0.09 with *T* more dominant for smaller *x* values.[10] Figure 2 plots the $c_0^f$-vs-*x* data. The solid and open symbols indicate *T*-LCCO and *T'*-LCCO, respectively. In the two-phase mixture region, the $c_0^f$ values for both *T*- and *T'*-LCCO are plotted. We observe a substantial monotonic shrinkage of $c_0^f$ with *x*, irrespective of *T*- or *T'*-LCCO, which confirms that Ce is actually incorporated into both the *T*- and *T'*-lattices under our film growth conditions.

Figure 3 provides the corresponding resistivity-vs-*x* plots, which show the resistivity value at 300 K [$\rho_{(300\ K)}$] as a function of Ce concentration for films grown on LAO substrates. The figure plots the data from both of the films cooled in a vacuum and those cooled in ozone. The former are denoted as "stoichiometric" since the vacuum-cooled films are almost completely oxygen-stoichiometric with presumably no excess oxygen,[11] whereas the latter are denoted as "oxidized" since ozone-cooled films have a fair amount of interstitial excess oxygen.[12] For the stoichiometric films, the Ce doping increases $\rho_{(300\ K)}$ significantly in the single-phase *T* region ($x \leq 0.045$). The Ce-doped films are highly insulating (~ $10^4$ Ωcm), and their $\rho_{(300\ K)}$ value is three to four orders of magnitude higher than that of pristine *T*-LCO. It should be noted that ~$10^4$ Ωcm is the upper limit of our resistivity measurements on 450 Å thick films, so the actual resistivity may be higher. Further Ce doping results in a slight inclusion of *T'*-phase materials, which greatly reduces $\rho_{(300\ K)}$ to $10^{-2}$ ~ $10^{-3}$ Ωcm. In the single-phase *T'* region ($x \geq 0.105$), $\rho_{(300\ K)}$ falls to 2 - 3 × $10^{-4}$ Ωcm. On the oxidized films, the introduction of excess oxygen has a totally opposite effect on the *T* and *T'* phases. Excess oxygen lowers the $\rho_{(300\ K)}$ value of the *T* films by four to seven orders of magnitude whereas it increases that of the *T'* films.

Figure 4 shows the superconducting transition temperature ($T_c$) of LCCO films as a function of Ce concentration for both stoichiometric and oxidized films grown on LAO substrates. Stoichiometric films with the *T*-structure are highly insulating and, unsurprisingly, not superconducting. The superconductivity appears only after the introduction of excess oxygen. The $T_c$ of oxidized *T*-LCCO films depends substantially on *x*, which is additional evidence supporting the notion that Ce is incorporated in the *T* lattice. If Ce *were* not in the *T* lattice, $T_c$ *would* be constant. The superconductivity observed in stoichiometric films at $x \geq 0.075$ is due to the co-existing *T'*-phase materials.[7, 8]

### B. Valence of Ce

Next we discuss the valence state of Ce dopant in the *T*-phase materials. It is known that Ce can be trivalent as well as tetravalent. If Ce *were* trivalent, it *would* not be surprising for *T*-LCCO to be insulating. It is difficult to evaluate the Ce valence of thin films by wet chemical analysis, so it was performed by *in-situ* X-ray photoelectron spectroscopy (XPS). Figure 5 shows the Ce 3*d* XP spectrum of *T*-$La_{1.94}Ce_{0.06}CuO_4$ film [(a)], together with previously reported reference spectra (Ref. 13) that are representative of the $Ce^{4+}$ and $Ce^{3+}$ states [(b) and (c)]. As shown in the reference spectra, the Ce 3*d* spectra are characterized by complex features that are related to the final state occupation of the Ce 4*f* level. The assignment of each peak structure is not always well-established but the consensus is that the highest binding energy peaks (U''' and V''') in spectrum (b) result from a Ce $3d^9 4f^0 O2p^6$ final state, which means this peak can be used as a hallmark of the tetravalent state.[13, 14] Although the resolution of spectrum (a) is fairly poor with a rather high background due to the low Ce concentration (about 0.86 at%) in *T*-$La_{1.94}Ce_{0.06}CuO_4$,[15] the spectrum shows distinct peaks at around 898 eV (V''') and 916 eV (U'''), indicating that the valence of Ce in the *T* phase is close to +4, as in the *T'* phase. This conclusion is also supported by the doping dependence of $c_0^f$ and by the doping dependence of $T_c$ in the oxidized films. As seen in Fig. 2, the slope of the $c_0^f$-$x$ plot is almost identical in *T*- and *T'*-LCCO,[16] indicating that the ionic radius of Ce in both compounds is close to $Ce^{4+}$ [0.97 Å for coordination number (CN) = 8 or 1.01 Å for CN = 9] rather than $Ce^{3+}$ [1.14 Å for CN = 8 or 1.196 Å for CN = 9].[17] Figure 6 compares the doping dependence of $T_c$ between oxidized *T*-$La_{2-x}Ce_xCuO_{4+\delta}$ (LCCO+) and *T*-LSCO.[18] Both of films are grown on LSAO substrates. As shown in Fig 6, slope of $T_c$ with $x$ are identical in the *T*-LCCO+ and the *T*-LSCO films, and the slope for the *T*-LCCO+ films is described by that for the *T*-LSCO films with simple shift, indicating that Ce substitution in the *T*-LCCO+ films compensates for hole doping with excess oxygen and valence of Ce is close to 4+ rather than 3+. Hence we conclude that electron doping is actually achieved with Ce doping.

Electron doping by $Ce^{4+}$ substitution in *T*-LCO has two effects on the lattice. One is the shrinkage of the $La_2O_2$ block layer resulting from the substitution of small $Ce^{4+}$ (1.01 Å) for large $La^{3+}$ (1.216 Å),[17] which reduces $c_0^f$ as seen in Fig. 2. The other is the enlargement of the $CuO_2$ plane caused by filling electrons in the Cu-O $dp\sigma_{x2-y2}$ anti-bonding band.[19] The two effects contribute to a decrease in the crystallographic tolerance factor, leading to the structural transition from *T* to *T'*.[20]

### C. Epitaxial stabilization of *T* phase

The structural transition from *T* to *T'* occurs at $x$ = 0.06 - 0.09 on LAO. If we are to expand the *T*-phase region to a higher $x$, the choice of substrate is important. Next we describe the epitaxial stabilization effect. Figure 7 shows the XRD patterns of LCCO films with the same $x$ (= 0.045) grown on different substrates. Caution is needed when identifying the phases of films on $K_2NiF_4$-type substrates since the peak positions for the films and substrates are close to each other because of the similarity of the crystal structures.[21] In particular, the peak positions of *T'*-LCCO for a certain Ce concentration range become very close to those of the substrates. As mentioned in Fig. 3, *T*-LCCO is highly insulating and a slight inclusion of *T'*-LCCO in *T*-LCCO greatly reduces $\rho_{(300\ K)}$. Hence, the results of $\rho_{(300\ K)}$-vs-$x$ plots are also considered to determine the phase of LCCO. It is safe to conclude that the film on LAO is single-phase *T*, the films on $KTaO_3$, $YAlO_3$, and $NdCaAlO_4$ (NCAO) are single-phase *T'*, and the film on $SrTiO_3$ is a mixture of *T* and *T'*. If we judge solely from XRD results, the films on $LaSrGaO_4$ (LSGO), $LaSrAlO_4$ (LSAO), $PrSrAlO_4$ (PSAO), and $NdSrAlO_4$ (NSAO) appear to be single-phase *T*. The films on LSGO and LSAO are highly insulating, so it is reasonably safe to conclude that these films are actually single-phase

*T*. In contrast, the films on PSAO and NSAO are fairly conductive, implying the inclusion of *T'*-phase materials. Hence we concluded that, for PSAO and NSAO, the peaks of the *T'*-LCCO films are hidden by those of the substrate.

## D. Phase diagram of *T* versus *T'* in $a_s$-$x$ plane

Identical investigations were performed for $x$ from 0 to 0.12 on all the substrates listed in Table I. The results are summarized in Fig. 8 as the phase diagram of *T* versus *T'* in the $a_s$-$x$ plane. Approximate boundaries are also indicated in the figure, which separate single-phase *T*, single-phase *T'*, and a two-phase mixture of *T* and *T'* regions. As $x$ increases there is a transition from *T* to *T'*. The critical Ce concentration ($x_{T\text{-}T'}$) for this transition is substrate-dependent. The $x_{T\text{-}T'}$ value is maximized at 0.075 - 0.12 with LSGO ($a_s$ = 3.843 Å). The $x_{T\text{-}T'}$ value gradually decreases for the $a_s$ value smaller than that of LSGO whereas it rapidly decreases for the $a_s$ value larger. The substrate dependence of $x_{T\text{-}T'}$ is explained by the epitaxial stabilization.[9] With no Ce doping, in-plane lattice constant of bulk *T*-LCO is ~ 3.803 Å, so substrates with $a_s$ ~ 3.80 Å such as LAO stabilize *T*-LCO. With Ce doping, the inherent value of the "bulk" in-plane lattice constant ($a_0^b$) of *T*-LCCO is unknown because bulk *T*-LCCO has not yet been synthesized. However, it should increase with increasing $x$ because the Cu-O bonds stretch with electron doping.[4, 19, 22] Hence, *T*-LCCO with a higher $x$ value is lattice-matched with substrates that have a correspondingly larger $a_s$. This explains qualitatively why $x_{T\text{-}T'}$ gradually increases with increasing $a_s$ for $a_s \leq 3.843$ Å (LSGO). A further increase in $a_s$ results in gradually better matching with *T'*-LCCO ($a_0^b$ ~ 4.01 Å) rather than *T*-LCCO, which explains the rapid drop of $x_{T\text{-}T'}$ from LSGO to KTO ($a_s$ = 3.989 Å) via STO ($a_s$ = 3.905

Å).

Figure 9 shows the temperature dependence of the resistivity for films grown on LSGO substrates with different $x$ values. As seen again in the figure, the $T$-LCCO films become more insulating with increasing $x$ up to 0.06. Metallic behavior with a superconducting transition is observed in films where $x \geq 0.075$, which is due to the inclusion of $T'$-phase materials.[7, 8]

### E. Epitaxial strain

Figure 10 compares the lattice constants, $a_0^{\mathrm{f}}$ and $c_0^{\mathrm{f}}$, of LCCO films on LSGO and LSAO substrates to illustrate the epitaxial strain effect. On an LSGO substrate, the $a_0^{\mathrm{f}}$ and $c_0^{\mathrm{f}}$ values of the film both show a rather weak dependence on $x$, whereas, on LSAO, $a_0^{\mathrm{f}}$ increases and $c_0^{\mathrm{f}}$ decreases monotonically with increasing $x$ up to the $T$-$T'$ phase boundary regime. The difference can be explained by taking the epitaxial strain effect into account. Although the precise values of the inherent lattice constants of "bulk" $T$-LCCO ($a_0^{\mathrm{b}}$ and $c_0^{\mathrm{b}}$) are unknown, the relation $a_{\mathrm{s}}$(LSAO) < $a_0^{\mathrm{b}}$($T$-LCCO) < $a_{\mathrm{s}}$(LSGO) undoubtedly holds for $x \leq 0.1$. Then $T$-LCCO films have in-plane compressive strain on LSAO whereas they have in-plane tensile strain on LSGO. As a result of the Poisson effect, the out-of-plane lattice constant expands on LSAO and shrinks on LSGO. Pristine $T$-LCO films on LSAO and LSGO are almost fully strained with $a_0^{\mathrm{f}}$ very close to $a_{\mathrm{s}}$. The $c_0^{\mathrm{f}}$ is 13.05 Å on LSGO and 13.22 Å on LSAO. With increasing $x$, $a_0^{\mathrm{b}}$ increases with electron doping, and thereby the lattice mismatch [$\equiv (a_{\mathrm{s}} - a_0^{\mathrm{b}}) / a_0^{\mathrm{b}}$] decreases between $T$-LCCO and LSGO, and increases between $T$-LCCO and LSAO. As seen in Fig. 10, the $a_0^{\mathrm{f}}$ of $T$-LCCO films on LSGO remains unchanged with doping and is almost equal to the $a_{\mathrm{s}}$ (3.843 Å) of LSGO, indicating that $T$-LCCO films on LSGO are fully strained for a whole range of $x$. In contrast, the $a_0^{\mathrm{f}}$ of $T$-LCCO films on LSAO increases steeply and approaches the $a_0^{\mathrm{f}}$ values on LSGO for higher $x$ values. This indicates that $T$-LCCO films on LSAO are fully strained only at $x = 0$, and become more strain-relaxed with increasing $x$. The different strain relaxation of the films on LSGO and LSAO is reflected also in the $c_0^{\mathrm{f}}$-vs-$x$ plot. The slope of $c_0^{\mathrm{f}}$-vs-$x$ is significantly different for the $T$-LCCO films on LSGO and on LSAO. The slope on LSAO is much steeper than that on LSGO. There are two contributors to the change in $c_0^{\mathrm{f}}$ with increasing $x$, namely the decrease due to $Ce^{4+}$ substitution and the change in the degree of strain. On LSAO, the compressive strain relaxes more as $x$ increases, which reduces the out-of-plane Poisson expansion, resulting in the additional decrease in $c_0^{\mathrm{f}}$. On LSGO, the tensile strain decreases because of the better lattice match with a higher $x$, which reduces the out-of-plane Poisson compression, resulting in partial compensation for the $c_0^{\mathrm{f}}$ decrease due to $Ce^{4+}$ substitution. The large epitaxial-strain effect observed in $T$-LCCO is hardly observed in $T$'-LCCO. In the $T$'-phase, the lattice mismatch between $T$'-LCCO ($a_0^{\mathrm{b}} \sim 4.01$ Å) and substrates ($a_{\mathrm{s}}$ = 3.843 Å for LSGO and 3.756 Å for LSAO) is too large to sustain epitaxial strain.

### F. Electron-hole symmetry

Figure 11 shows "global" electronic phase diagrams from hole to electron doping in $T$-LCO, where (a) $T_{\mathrm{c}}$ and (b) $\rho_{(300\ \mathrm{K})}$ are plotted as a function of Ce or Sr concentration. The data for the electron-doped regime are taken from the present work on $T$-LCCO grown on LSGO, in which the electron doping is maximized up to $x = 0.06$. The data for the hole-doped regime are taken from our previous work on $T$-LSCO grown on LSAO.[18, 23] In $T$-LCO, superconductivity appears only with hole doping although it cannot be ruled out that superconductivity might appear with electron doping beyond 0.06. More important is the behavior of $\rho_{(300\ \mathrm{K})}$. Electron doping and hole doping in *T*-LCO provide a sharp contrast. Hole doping metallizes *T*-LCO whereas electron doping makes *T*-LCO more insulating. The $\rho_{(300\ \mathrm{K})}$ value starts to rise rapidly with a break in the slope of the $\rho_{(300\ \mathrm{K})}$-versus-$x$ curve for hole doping of about 0.05 and steadily increases up to electron doping of about 0.06. In fact, we do not see any singularity at $x = 0$. Figure 11 appears to show broken electron-hole symmetry in the *T* phase.

As mentioned in the *Introduction*, it has been claimed that the electron-hole symmetry appears to hold in high-$T_c$ superconductors, but this argument is based on a comparison of hole and electron doping in different structures, namely, hole doping in the $K_2NiF_4$ (*T*) structure (Ref. 3) and electron doping in the $Nd_2CuO_4$ (*T'*) structure.[4] It has to be emphasized that the mother compounds of the two structures, *T*-LCO and *T'*-LCO, have quite different electronic properties as revealed by our previous work.[9] *T*-LCO is highly insulating whereas *T'*-LCO is fairly metallic.[24] Furthermore a striking contrast is observed for Ce doping in *T*-LCO and *T'*-LCO: Ce doping makes *T*-LCO insulating whereas it makes *T'*-LCO superconducting (Fig. 9). This indicates that the Cu-O coordination totally changes the electronic phase diagram of HTSC. Hence it is desirable to establish a global electronic phase diagram from hole to electron doping in the same crystal structure with octahedral, pyramidal, and square-planar Cu-O coordinations. Such attempts have been made with respect to bulk synthesis but have proved unsuccessful. Our present work, however, demonstrates that a low-temperature thin-film synthesis route may enable such an exploration.

### G. Nature of insulating state in *T*-LCCO

Next, we must investigate the nature of the insulating state in electron-doped *T*-LCO. If *T*-LCO *were* a simple Mott insulator, electron doping *would* lead to a metallic state. But our experimental results show just the opposite. Then, we have to think of other possible insulating mechanisms in electron-doped *T*-LCO. One simple explanation is that electron doping by Ce compensates for hole doping with remnant excess oxygen, which is not removed when films are cooled in a vacuum. However, if it *were* the case, as much as $\delta = 0.03$ excess oxygen *would* be required after vacuum cooling to account for the fact that *T*-LCCO films are still insulating with Ce doping up to $x = 0.06$. This is unlikely.

Another explanation is to regard the antiferromagnetic (AF) insulating ground state in *T*-LCO as a Fermi-surface driven spin-density-wave (SDW) state. Then, Fermi surface nesting could occur over a larger area at finite electron doping than at zero doping, depending on the shape of the Fermi surface. However, the Fermi surface nesting is usually incomplete in two- or three-dimensional materials and leaves some pieces of the Fermi surface, hence, it may be hard to explain the highly insulating state of *T*-LCO.

The third explanation is based on the ionic model, which describes *T*-LCO as a charge-transfer insulator. In this framework, doped holes go on the oxygen sites, creating spin frustration. The long-range AF order is promptly destroyed with hole doping, thereby leading to metallic conduction and superconductivity. In contrast, doped electrons go on the Cu sites, eliminating Cu spins leading to spin dilution instead of spin frustration. The AF order may not be destroyed until deep electron doping. In this AF lattice with $Cu^{2+}$ ($S = 1/2$, $d^9$), even though electrons are doped, they are not easy to move. For example, an up-spin electron that comes on a down-spin Cu site is unable to hop to the nearest neighbor up-spin Cu sites as described by the Pauli principle. It can hop only to the Cu 4*s* level at the cost of the 3*d*-4*s* energy difference.

As the final possible explanation, we should note that the insulating behavior starts at hole doping of ~ 0.05 and is continuous at $x = 0$. So the insulating nature in the hole-doped and electron-doped regimes may have the same origin. In hole-doped *T*-LSCO, the low-temperature upturn in normal-state resistivity exists even at optimum doping, develops as $x$ decreases, and rapidly grows below $x = 0.05$ with the disappearance of superconductivity.[25, 26] There have been certain indications that the development of the low-temperature upturn is related to the growth of the pseudo-gap. We have proposed the Kondo effect as the origin of the low-temperature resistivity upturn near the optimum doping.[25, 26] This naturally leads us to regard *T*-LSCO with $x < 0.05$ as a Kondo insulator. In this scenario, strong hybridization between O2*p* holes and Cu3*d* electrons leads to a large Kondo coupling ($J_K$). Hence, each O2*p* hole is strongly bound to Cu3*d* spins, eventually forming a localized Kondo singlet (Zhang-Rice singlet) with a small energy gap. Of course, this picture cannot simply be extended to the electron-doped regime or even to the region where the long-range AF order develops since the AF order freezes the flipping of local spins that is indispensable to the Kondo effect. At present we speculate that the Kondo interaction ($J_K$) and the exchange interaction ($J$) are two essential ingredients to an understanding of the very complicated nature of the insulating state in lightly electron-doped and hole-doped *T*-LCO.[25, 26]

## IV. SUMMARY

We synthesized $T$-$La_{2-x}Ce_xCuO_4$ with $x \leq 0.06$ by molecular beam epitaxy and investigated the effect of electron doping on $T$-$La_2CuO_4$. The key to synthesizing $T$-$La_{2-x}Ce_xCuO_4$ with large $x$ is the choice of lattice-matched substrate, $LaSrGaO_4$ ($a_s$ = 3.843 Å). All $T$-$La_{2-x}Ce_xCuO_4$ films were insulating and became more insulating with electron doping. The result indicates that the electron-hole symmetry is broken in $T$-$La_2CuO_4$. The insulating nature in $T$-$La_{2-x}Ce_xCuO_4$ is ascribed to not a Mott insulator but a Kondo insulator.

## ACKNOWLEDGMENT

The authors thank Dr. S. Karimoto, and Dr. H. Shibata, Dr. H. Sato, Dr. K. Ueda, Dr. T. Makimoto, Dr. T. Yamada, and Dr. A. Matsuda for helpful discussions, and Dr. K. Torimitsu, Dr. M. Morita, Dr. H. Takayanagi, and Dr. S. Ishihara for their support and encouragement throughout the course of this study.

## Figure captions

FIG. 1. X-ray diffraction patterns of $La_{2-x}Ce_xCuO_4$ films grown on $LaAlO_3$ (LAO) substrates with different Ce concentration *x*. The top two patterns are simulations for *T*- and *T'*-$La_2CuO_4$. The broken and dotted lines indicate the positions of the 008 diffraction peaks for *T*- and *T'*-$La_2CuO_4$, respectively. Peak positions of LAO substrates are indicated in the lowest figure.

FIG. 2 Variation in the *c*-axis lattice constant ($c_0^{\mathrm{f}}$) as a function of Ce concentration (*x*) for $La_{2-x}Ce_xCuO_4$ (LCCO) thin films grown on $LaAlO_3$ substrates. The filled and open circles indicate *T*- and *T'*-LCCO films, respectively. The dotted lines are a guide for the eyes.

FIG. 3. Plot of the resistivity at 300 K [$\rho_{(300\ \mathrm{K})}$] for $La_{2-x}Ce_xCuO_{4+\delta}$ films grown on $LaAlO_3$ substrates with various Ce concentrations (*x*). The filled, open, and gray circles indicate stoichiometric films ($\delta \sim 0$) with a single phase *T*, single phase *T'*, and a mixed phase of *T* and *T'*, respectively. The crosses, squares, and crosses with squares indicate oxidized films ($\delta > 0$) with a single phase *T*, single phase *T'*, and a mixed *T* and *T'* phase , respectively. Each phase regime is indicated at the top of the figure.

FIG. 4. Ce concentration (*x*) dependence of $T_c$ for $La_{2-x}Ce_xCuO_{4+\delta}$ films grown on $LaAlO_3$ substrates. The filled, open, and gray circles indicate stoichiometric films ($\delta \sim 0$) with a single phase *T*, single phase *T'*, and a mixed *T* and *T'* phase, respectively. The crosses, squares, and crosses with squares indicate oxidized films ($\delta > 0$) with a single phase *T*, single phase *T'*, and a mixed *T* and *T'* phase, respectively. Each phase regime is indicated at the top of the figure.

FIG. 5. (a) *In-situ* Ce 3$d$ XP spectrum of $T$-$La_{1.94}Ce_{0.06}CuO_4$ thin film. Spectra (b) $CeO_2$ ($Ce^{4+}$) and (c) Ce(III) oxide ($Ce^{3+}$) are for comparison (Ref. 13). Peaks U, U'', and U''' originate from Ce $3d_{3/2}$ whereas peaks V, V'', and V''' originate from Ce $3d_{5/2}$.

FIG. 6. Variation in $T_c$ for $La_{2-x}Ce_xCuO_{4+\delta}$ and $T$-$La_{2-x}Sr_xCuO_4$ (○) (Refs. 18, 23) as a function of Sr or Ce concentration ($x$). The crosses, squares, and crosses with squares indicate oxidized films ($\delta > 0$) with a single phase $T$, single phase $T'$, and a mixed $T$ and $T'$ phase , respectively.

FIG. 7. X-ray diffraction patterns of $La_{2-x}Ce_xCuO_4$ films grown on various substrates with $x = 0.045$. The top two patterns are simulations for $T$- and $T'$-$La_2CuO_4$. The broken and dotted lines indicate the positions of the 008 diffraction peaks for $T$- and $T'$-$La_2CuO_4$, respectively. Asterisks (*) indicate the substrate peaks.

FIG. 8. Phase diagram of the selective stabilization of $T$ versus $T'$ in the $a_s$-$x$ plane. The filled, open, and gray circles are for films with a single phase $T$, single phase $T'$, and a mixed $T$ and $T'$ phase, respectively. Each region is roughly separated by broken lines. The in-plane lattice constant of each substrate is indicated on the right axis: $K_2NiF_4$ type (♦) and perovskite (□) substrates.

FIG. 9. Temperature dependence of resistivity for $La_{2-x}Ce_xCuO_4$ films grown on $LaSrGaO_4$ substrates with different $x$ values. The solid lines indicate single-phase $T$ films (●: $x = 0$, ■: 0.015, ♦: 0.03, ▲: 0.045, ×: 0.06), while the broken line indicates mixed $T$- and $T'$-phase films (○: 0.075, □: 0.09, ◊: 0.105, +: 0.12).

FIG. 10. Variation in the $c$- and $a$-axis lattice constant ($c_0^f$ and $a_0^f$) as a function of Ce concentration ($x$) for $La_{2-x}Ce_xCuO_4$ thin films grown on $LaSrGaO_4$ (LSGO) and $LaSrAlO_4$ (LSAO) substrates. The circles and squares indicate films grown on LSGO and LSAO substrates, respectively. The filled, open, and gray symbols indicate films with a single phase *T*, single phase *T'*, and a mixed *T* and *T'* phase, respectively. The dotted lines are a guide for the eyes. The broken lines represent $a_s$ = 3.843 Å for LSGO and 3.755 Å for LSAO substrates.

FIG. 11. Variation in (a) $T_c$ and (b) resistivity at 300 K [$\rho_{(300\ K)}$] for *T*-$La_{2-x}Ce_xCuO_4$ (●) and *T*-$La_{2-x}Sr_xCuO_4$ (○) (Refs. 18, 23) as a function of Sr or Ce concentration ($x$).

Table I. The $a$-axis lattice constants ($a_s$) for the substrates used in this work.

| Substrate | Abbreviation | $a_s$ (Å) | Crystal structure |
|---|---|---|---|
| $KTaO_3$ | KTO | 3.989 | Perovskite |
| $SrTiO_3$ | STO | 3.905 | Perovskite |
| $LaSrGaO_4$ | LSGO | 3.843 | $K_2NiF_4$ |
| $LaAlO_3$ | LAO | 3.793 | Perovskite |
| $LaSrAlO_4$ | LSAO | 3.755 | $K_2NiF_4$ |
| $PrSrAlO_4$ | PSAO | 3.727 | $K_2NiF_4$ |
| $YAlO_3$ | YAO | 3.715 | Perovskite |
| $NdSrAlO_4$ | NSAO | 3.712 | $K_2NiF_4$ |
| $NdCaAlO_4$ | NCAO | 3.688 | $K_2NiF_4$ |

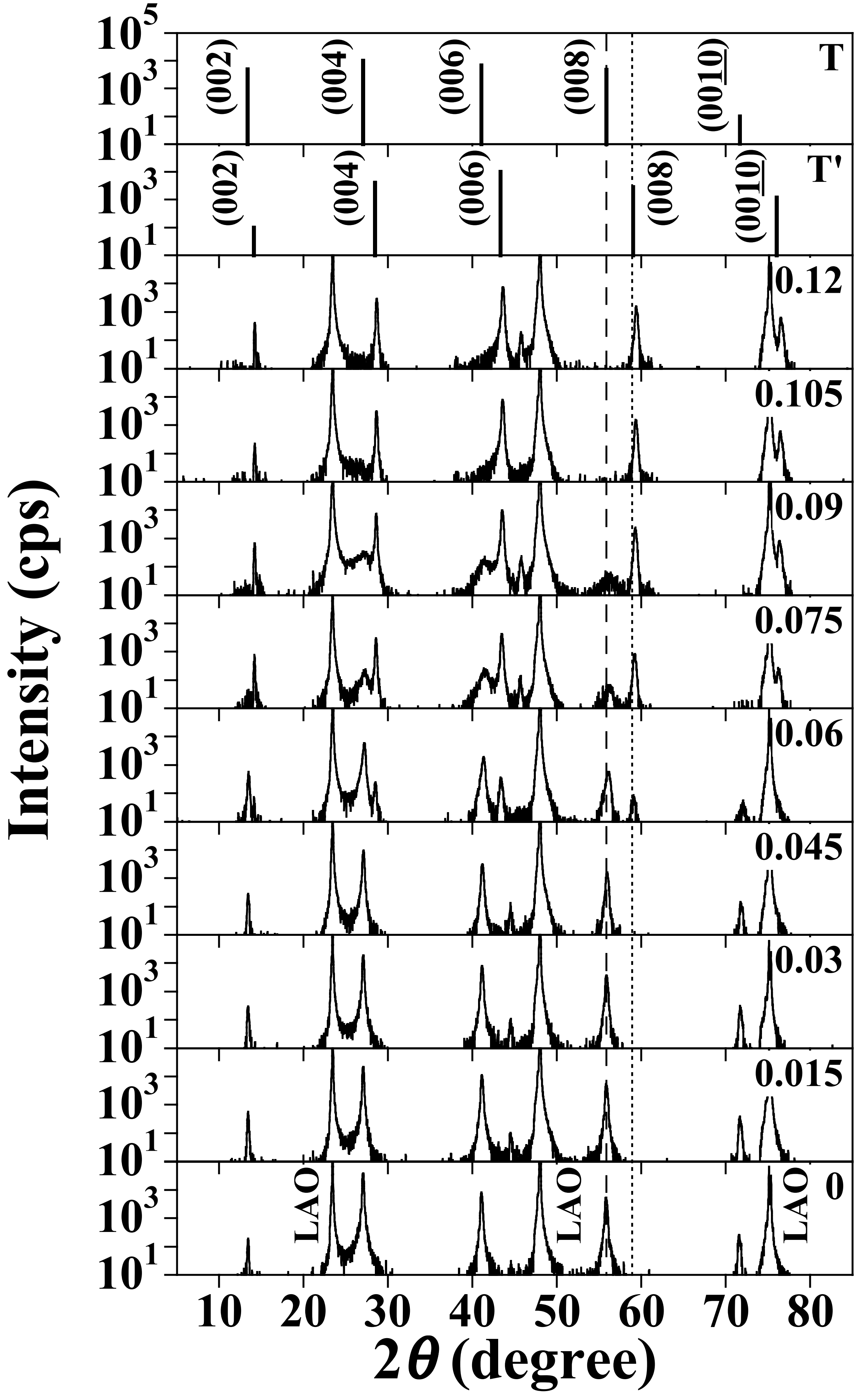

**Fig. 1, A. Tsukada *et al.***

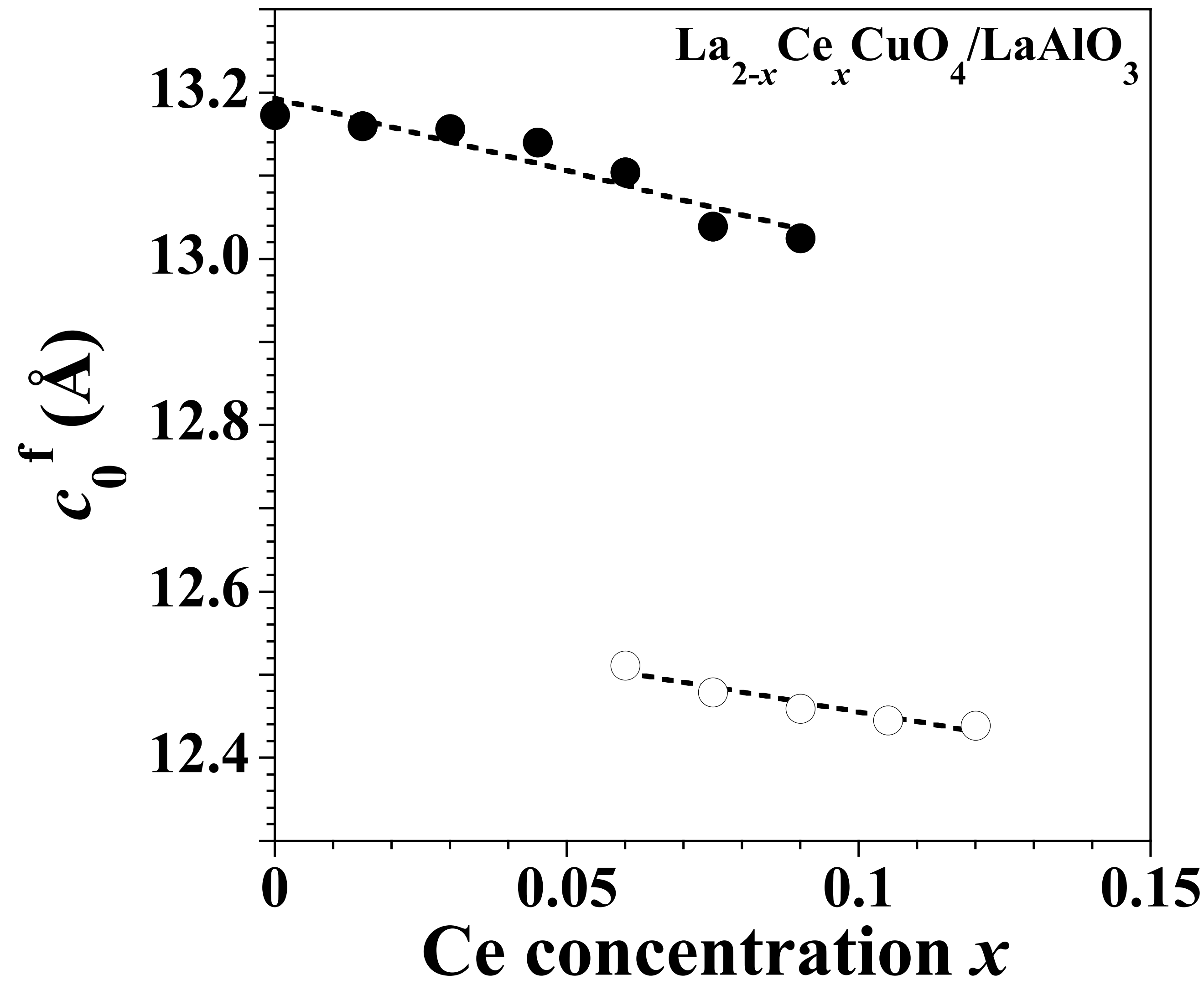

**Fig. 2, A. Tsukada *et al.***

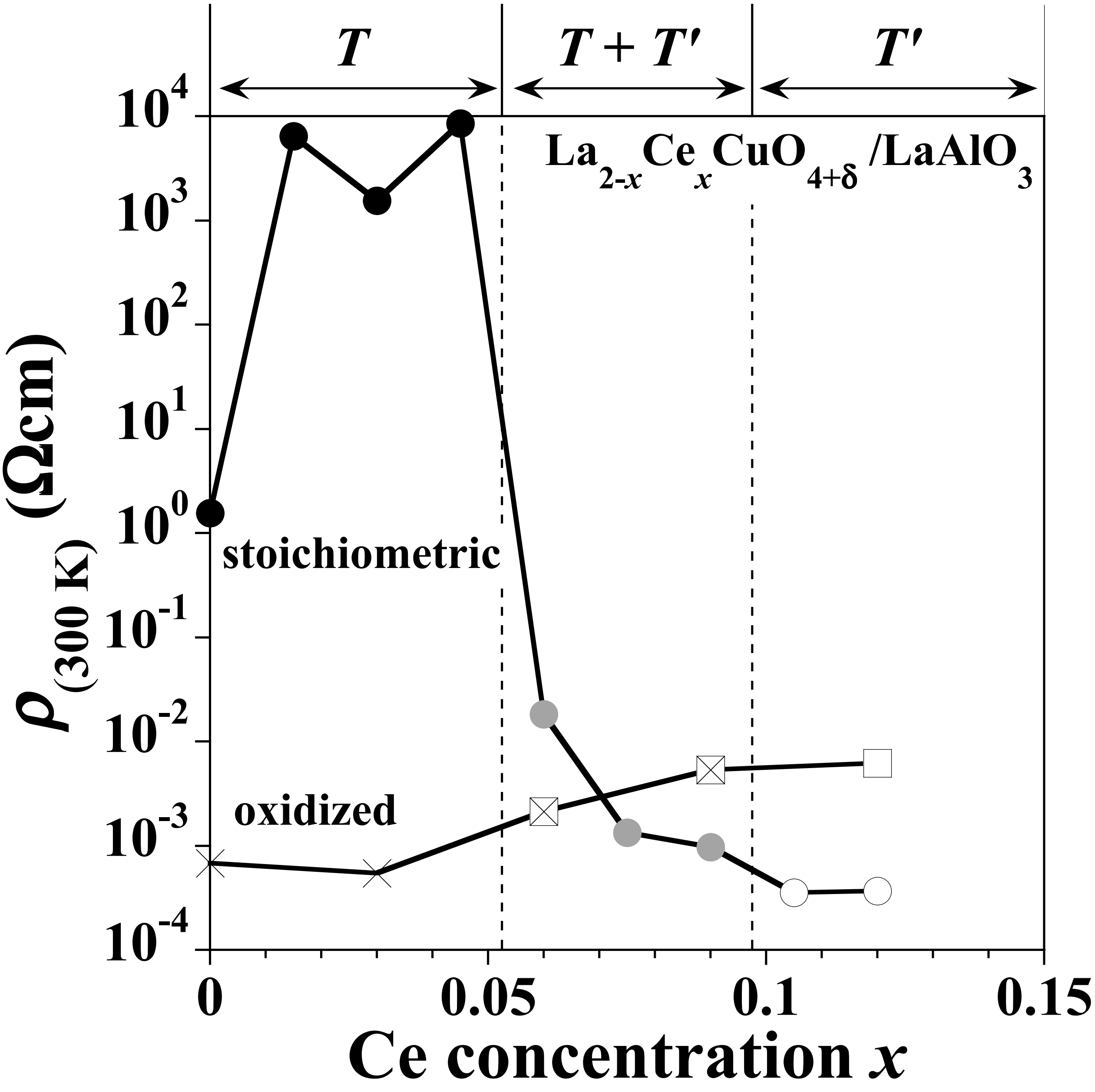

Fig. 3, A. Tsukada *et al.*

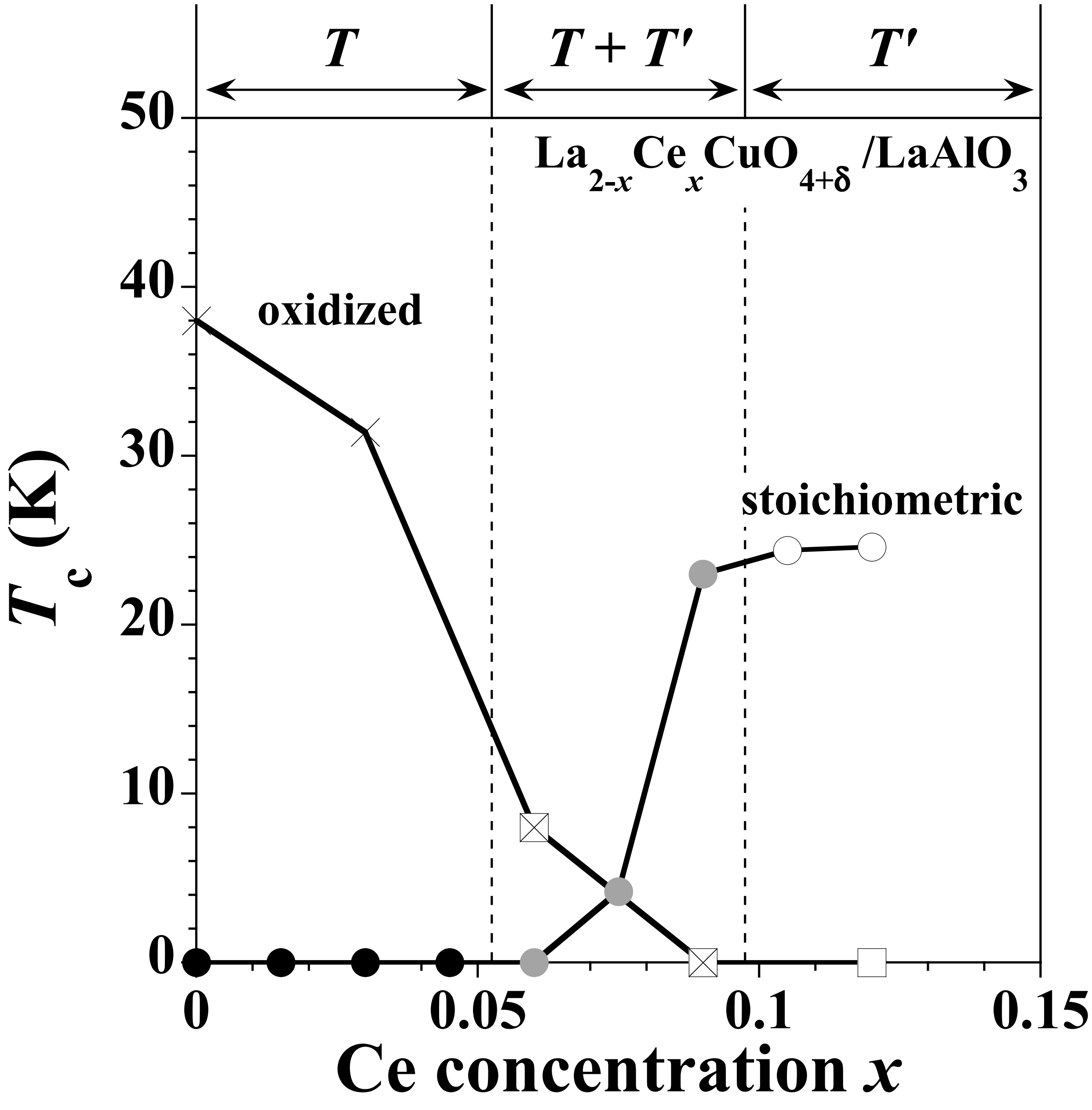

Fig. 4, A. Tsukada *et al.*

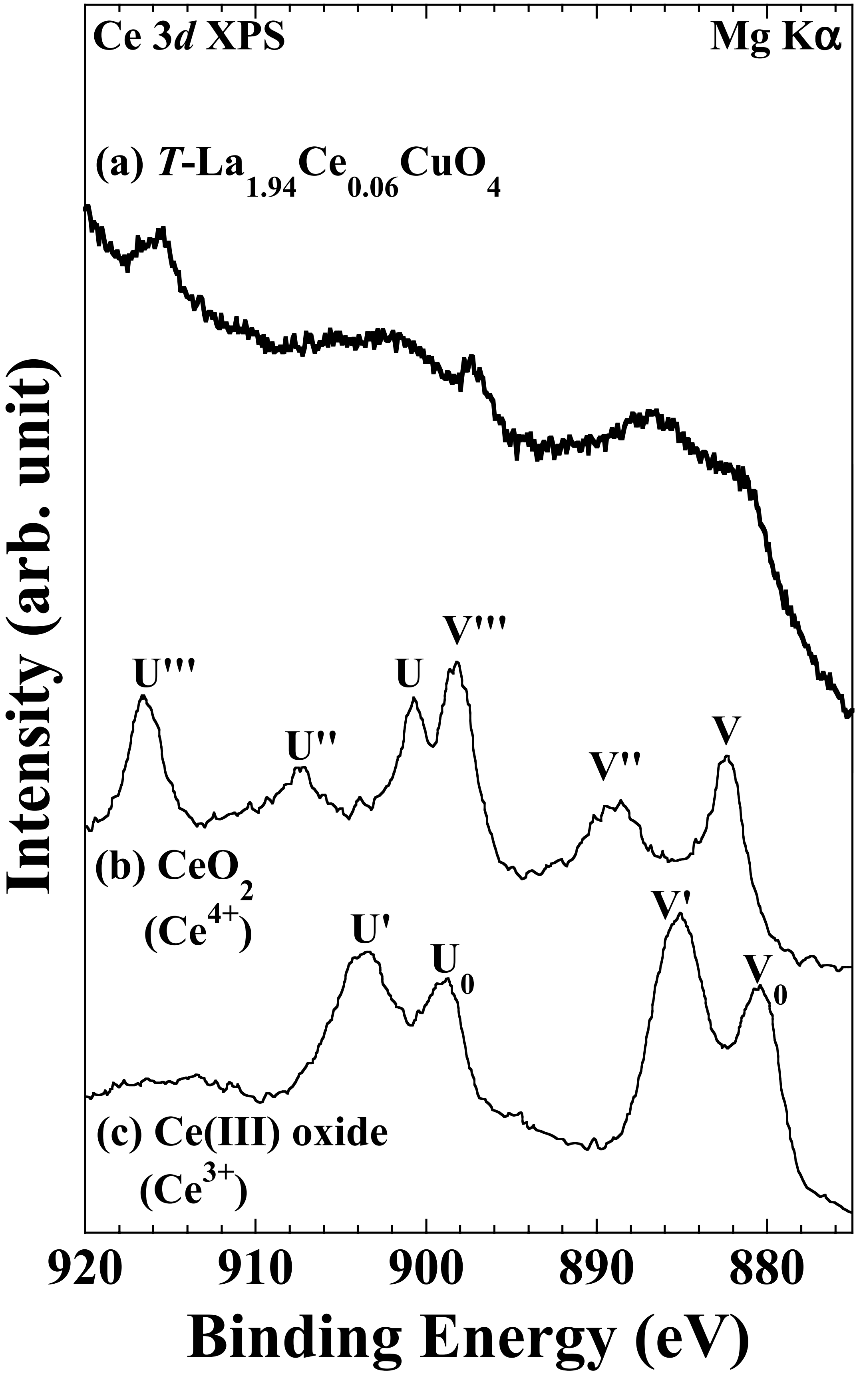

**Fig. 5, A. Tsukada *et al.***

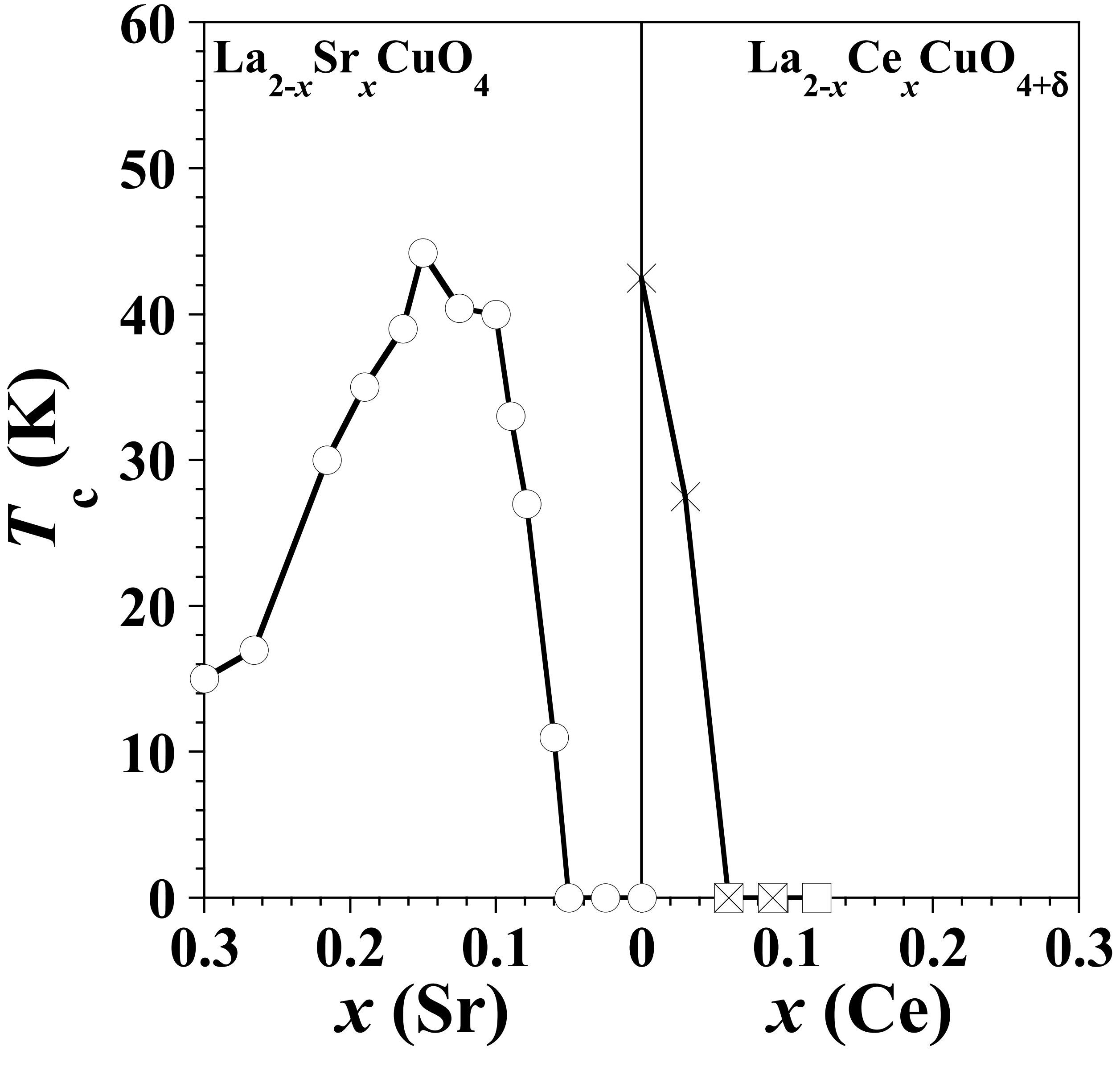

**Fig. 6, A. Tsukada *et al.***

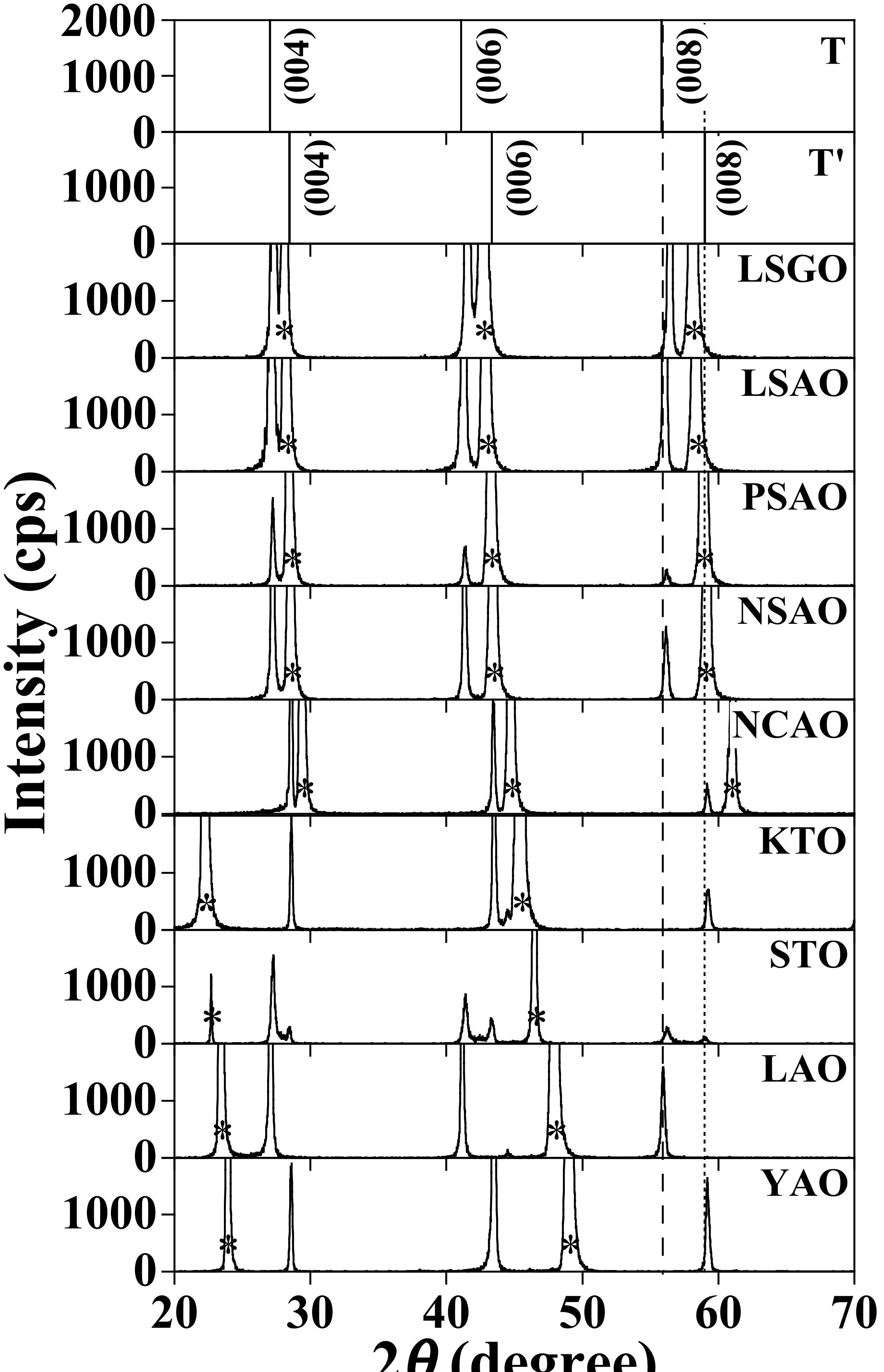

Fig. 7, A. Tsukada *et al.*

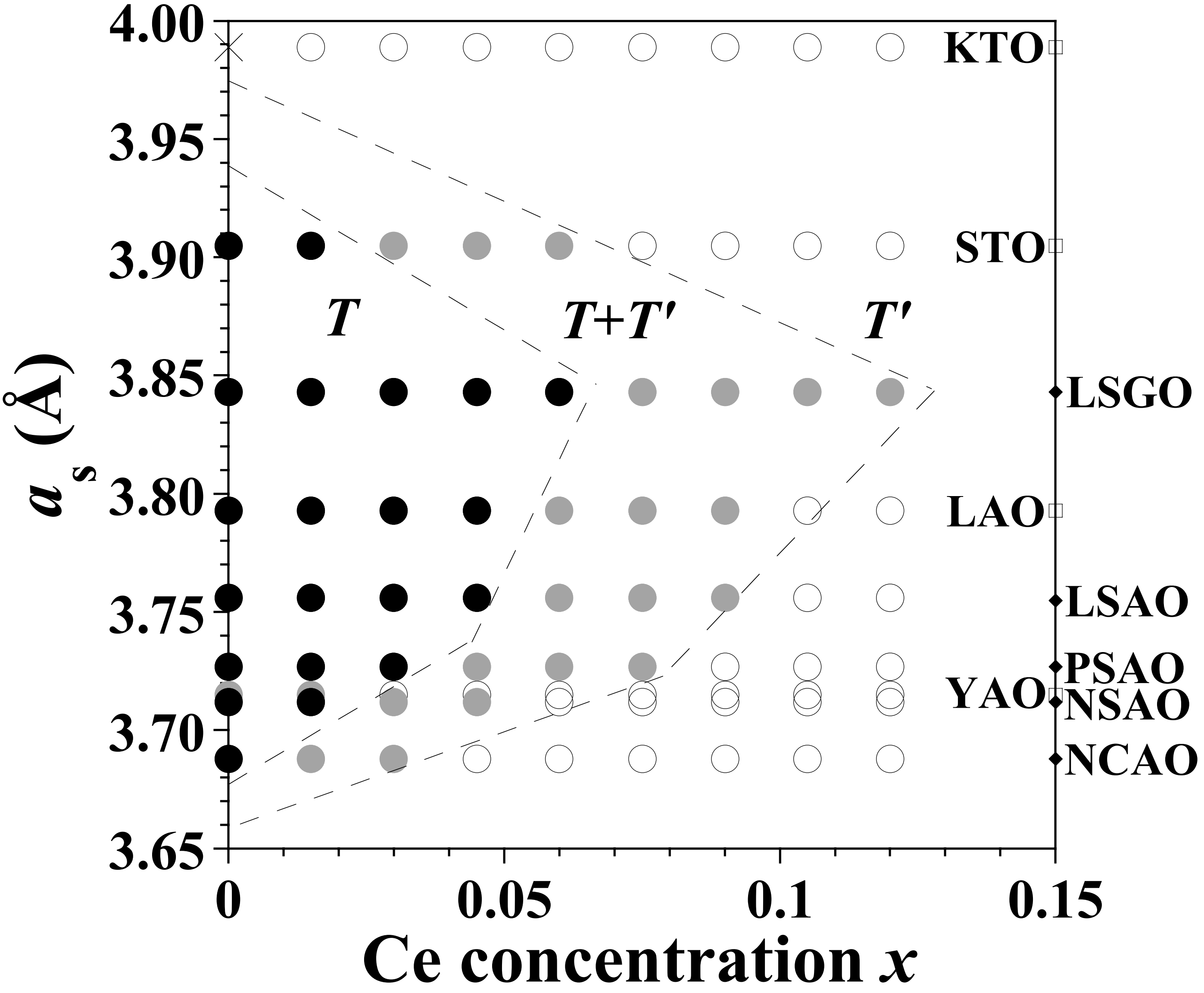

**Fig. 8, A. Tsukada *et al.***

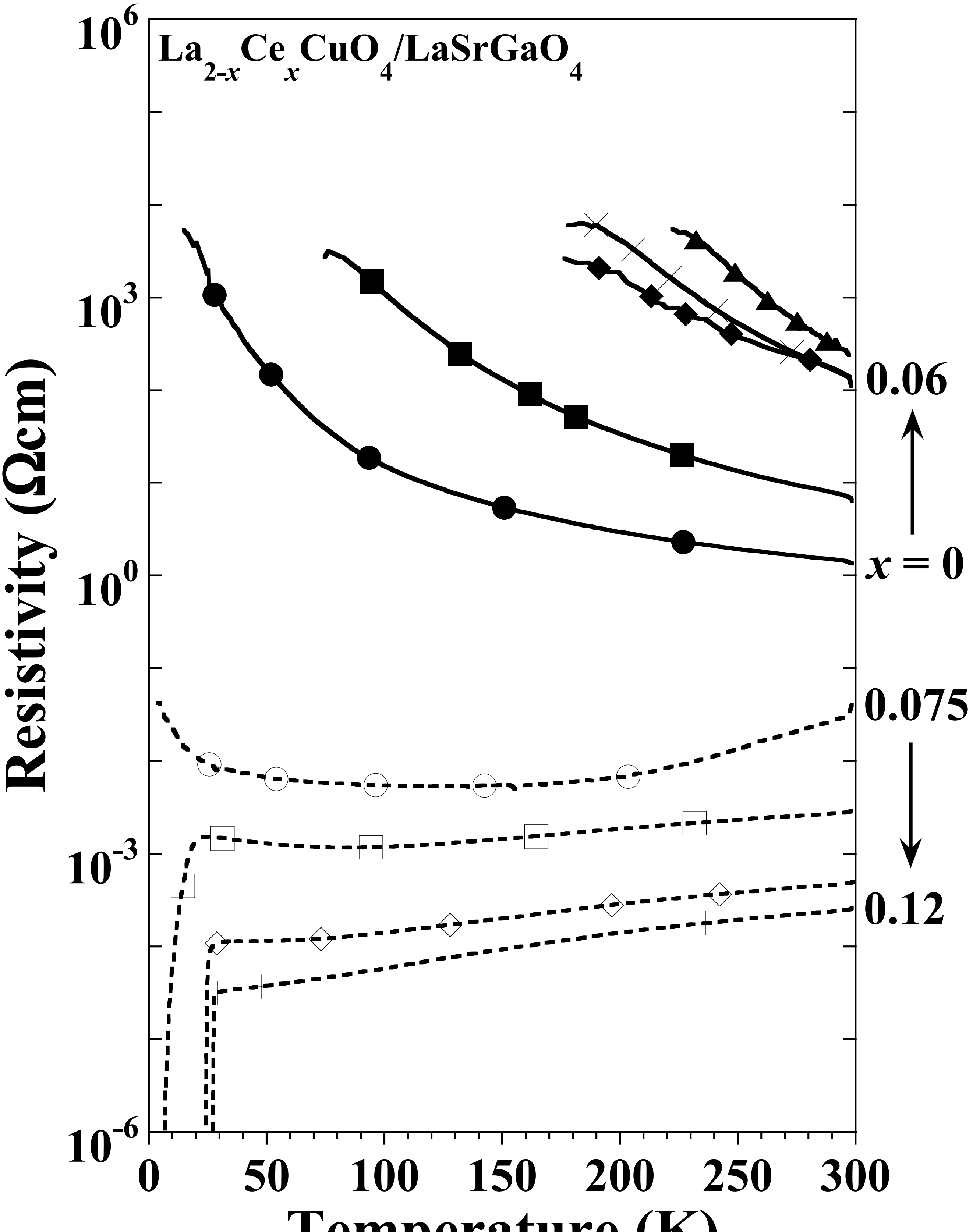

Fig. 9, A. Tsukada *et al.*

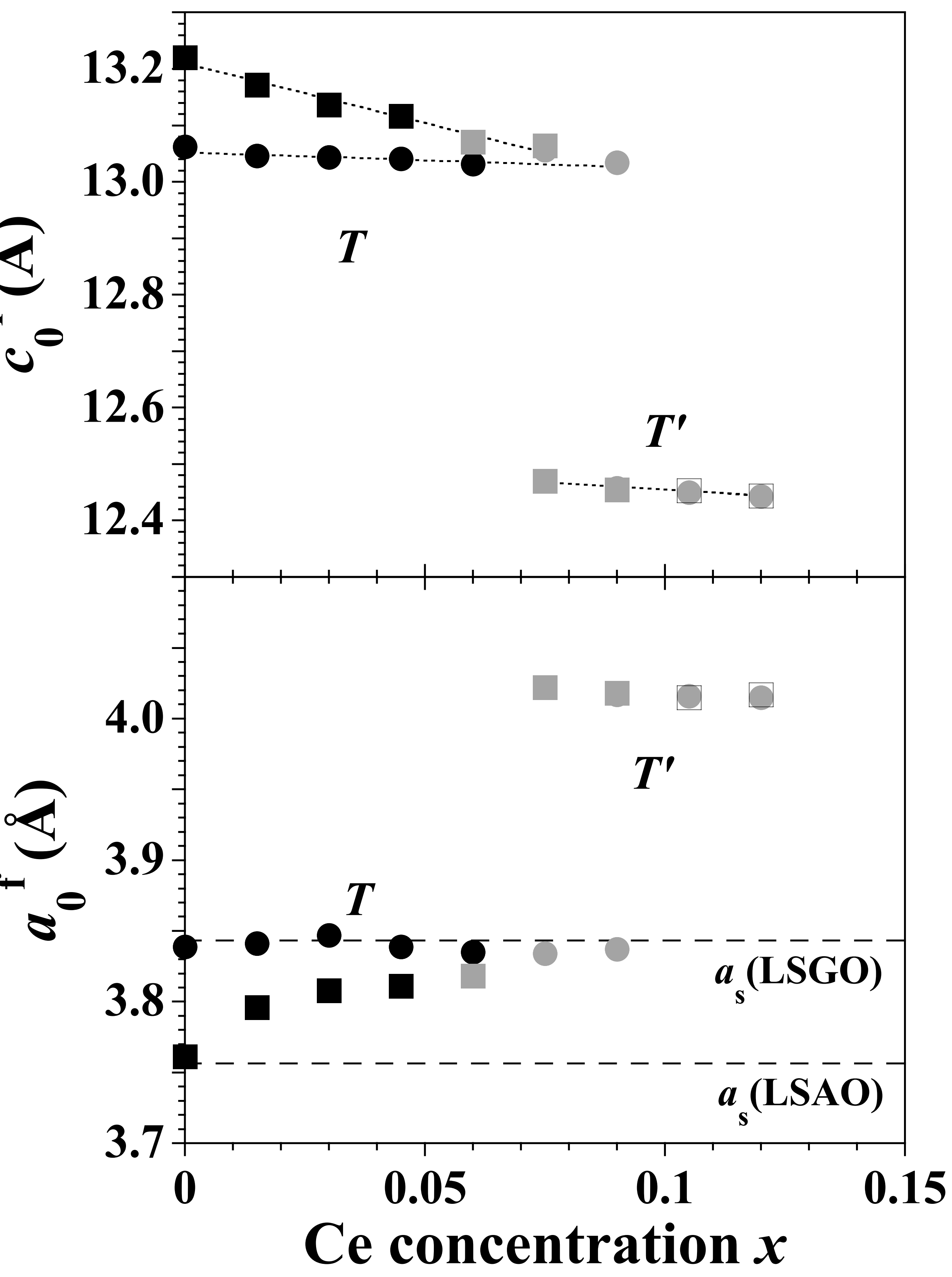

**Fig. 10, A. Tsukada *et al.***

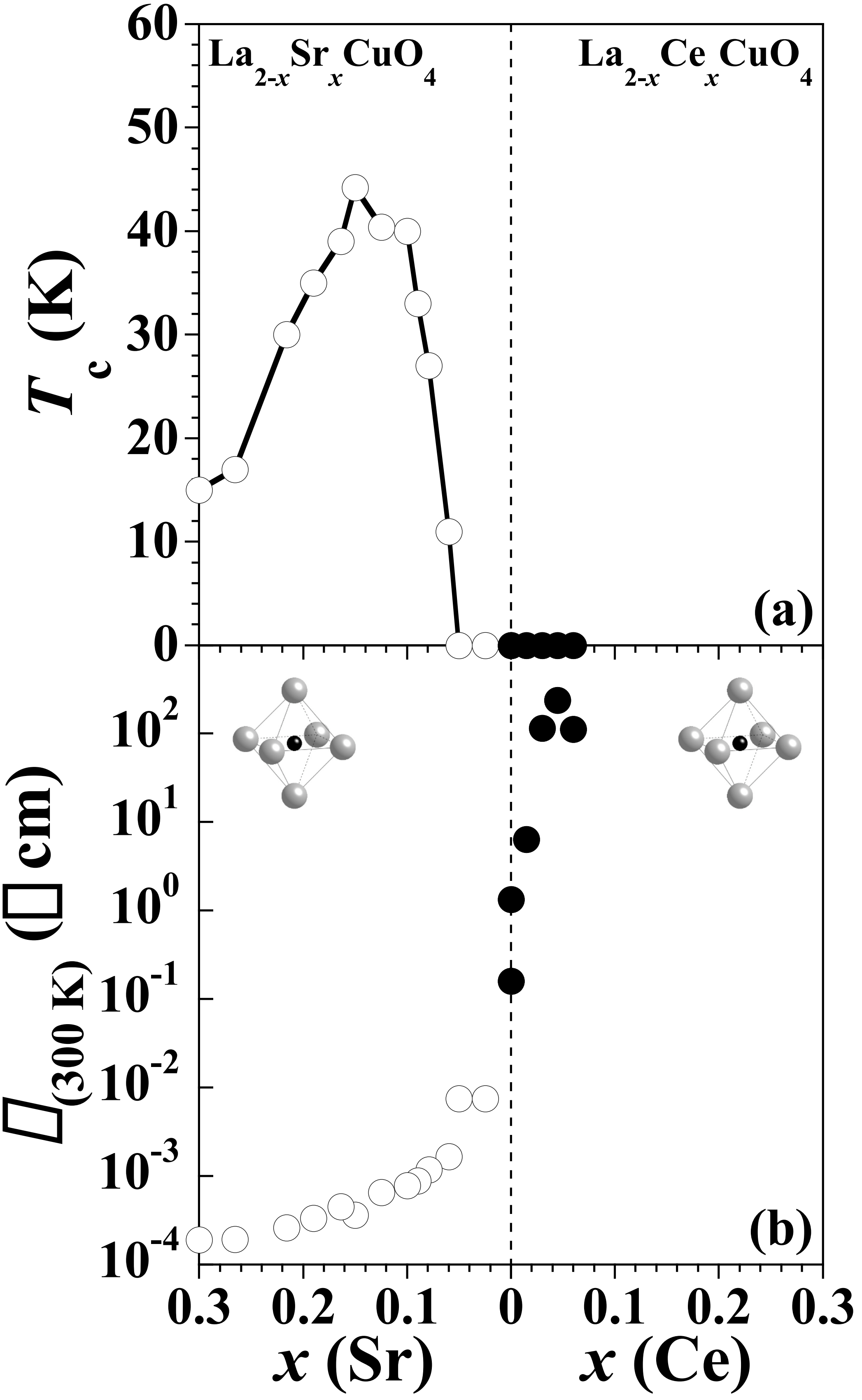

Fig. 11, A. Tsukada *et al.*